\documentclass[aps,twocolumn,prb,showpacs,eqsecnum]{revtex4-1}
\usepackage{graphpap}
\usepackage[dvips]{graphicx}
\usepackage[dvips]{graphics}
\usepackage{color}

\newcommand{\rhoF}{\rho_{\mathrm{f}}}
\newcommand{\EF}{E_{\mathrm{F}}}
\newcommand{\rr}{\mathbf{r}}
\newcommand{\ddd}{d}
\newcommand{\EFr}{E_{\mathrm{F}}(\rr)}

\begin{document}


\title{Negative quantum capacitance in graphene nanoribbons \\ with lateral gates} 

 \author{R. Reiter$^{1,*}$, U. Derra$^{2}$, S. Birner$^{3}$, B. Terr\'es$^{2}$, F. Libisch$^{1,4}$, J. Burgd\"orfer$^1$ and C. Stampfer$^{2}$}
 \affiliation{
$^1$\,Institute for Theoretical Physics, Vienna University of Technology, 1040 Wien, Austria, EU\\
$^2$\,JARA-FIT and II. Institute of Physics, RWTH Aachen University, 52074 Aachen, Germany, EU\\
$^3$\,Walter Schottky Institute and Institute for Nanoelectronics, 
  TU Munich, 85748 Garching, Germany, EU\\
$^4$\,Department of Mechanical and Aerospace Engineering, 
  Princeton University, Princeton, NJ, USA}





\date{\today}

\begin{abstract}

We present numerical simulations of the capacitive coupling between
graphene nanoribbons of various widths and gate electrodes in
different configurations.  We compare the influence of lateral
metallic or graphene side gate structures on the overall back gate
capacitive coupling. Most interestingly, we find a complex interplay
between quantum capacitance effects in the graphene nanoribbon and the
lateral graphene side gates, giving rise to an unconventional negative
quantum capacitance.  The emerging non-linear capacitive couplings are
investigated in detail. The experimentally relevant relative lever
arm, the ratio between the coupling of the different
gate structures, is discussed.
\end{abstract}


\maketitle 

\section{Introduction}

Graphene nanoribbons~\cite{che07,han07,li08,tod09,sta09,jia10,cai10} 
 are of  increasing interest due to their
promise of a band gap, overcoming the gapless
band structure of truly two-dimensional (2D) graphene~\cite{kat12,guinea_review}. 
%
In particular, their overall semiconducting behavior allows the
fabrication of graphene field-effect transistors~\cite{wan08},
tunnelling barriers~\cite{gue11}, and quantum
devices~\cite{ihn10}. First experimental demonstrations of graphene
nanoribbon based quantum dots~\cite{liu09,wan13}, double quantum
dots~\cite{liu10} and charge sensors~\cite{gue08, neu13} have been
reported in recent years.  In most of these quantum devices the local
electrostatic tunability of the electrochemical potential along
graphene nanoribbons is key for the device functionality. For this
purpose, local top-gates and lateral gates, based either on metals or
in-plane graphene have been fabricated (see,
e.g., Refs.~\onlinecite{Meric08, Williams07, Young09, Stander09}). In particular,
the 2D nature of graphene makes it straightforward to pattern a number of lateral
graphene gates and in-plane charge detectors from the very same
graphene sheet as the adjacent top-down fabricated
nanoribbon~\cite{ihn10,mol10}.  Consequently, a better understanding
of the capacitive coupling between nanoribbons (with different widths)
and (graphene) gate electrodes is important for device optimization
and future graphene-based nanoelectronics.  This is particularly true
for graphene nanodevices since their Fermi energy is tuned using
capacitive coupling. The strength of these couplings, in turn, depends
on the density of states (DOS) via quantum capacitance
effects~\cite{lur88,john04}. The low DOS close to the Dirac
point thus makes graphene a rather unusual gate material with reduced
and energy dependent screening properties~\cite{Fogler07,Ghaznavi10}.  
Graphene sheets have been subject to a large number of theoretical and
experimental studies of quantum capacitance
effects~\cite{xia09,xia10,xu11,tahir12,parrish12,kliros12,giannazzo09,citeulike:11024917,dro10,Ilani06,Wang13}. Indeed,
recent advances allowed experiments based on capacitance measurements
to observe such phenomena as Fermi velocity
renormalization~\cite{yu13}, fractional quantum Hall phase
transitions~\cite{fel13}, and Hofstadter's
butterfly~\cite{hun13}. Furthermore, quantum Hall transport has been
used to probe the capacitance profile at graphene
edges~\cite{ver13}. These demonstrations give vested hope that quantum
capacitance effects in graphene nano\-devices could be exploited in
future applications.

\begin{figure}
\includegraphics[draft=false,keepaspectratio=true,clip]{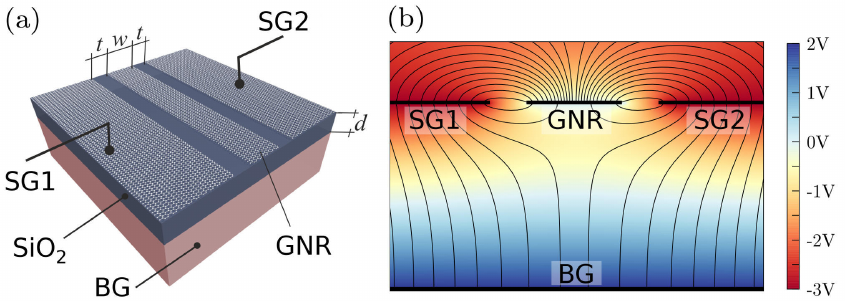}%
\caption{\label{system-illustration}(a) Schematic illustration of a typical
  model system: a graphene nanoribbon (GNR) on an isolating
  substrate (dark grey), capacitively coupled to a back gate (BG, light
  red) and side gates (SG1, SG2). $t$=30~nm, $d$=300~nm and varying
  $w$.  (b) Electrostatic potential and field lines on a two-dimensional
  plane perpendicular to the nanoribbon direction. Voltages are
  $V_{\mathrm{BG}}=2\,$V and $V_{\mathrm{SG}}=-3\,$V, nanoribbon width $w$ = 90~nm.}%
\end{figure}

  Here we present numerical simulations of the electrostatic
  capacitive coupling between graphene nanoribbons of different widths
  and gates in different configurations. In particular we elucidate
  the influence of metallic and graphene lateral side gate
  structures (see Fig.~\ref{system-illustration}) on the back gate
  capacitive coupling. We show that the non-linear capacitive
  couplings give rise to complex interplay between quantum capacitance
  effects in the nanoribbon and the gates. Surprisingly, we find the
  unconventional phenomenon of an effective negative quantum
  capacitance in nanoribbons with lateral graphene gates. Inclusion
  of quantum capacitance effects will therefore be
  essential in reliably interpreting experimental results.

The paper is organized as follows: in Sec.~\ref{sec:qc} we briefly
introduce the description of quantum capacitance effects on the
Thomas-Fermi level and the model system under investigation. Results
for graphene nanoribbons with different width, and with metallic and
graphene side gates are discussed in Sec.~\ref{sec:results}. Finally,
we compare the capa\-citive coupling between either nanoribbon and
back gate or nanoribbon and side gates.  We conclude with a short summary.

\section{Quantum capacitance and model system}\label{sec:qc}

Accurate simulations of the influence of side gates require a detailed
treatment of the electrostatic problem beyond analytical models available for
simple geometries such as the infinitely extended graphene sheet or an isolated
nanoribbon.\cite{lin08,citeulike:11024917} We therefore present a
self-contained treatment of the quantum capacitance effect, and
present a numerical approach to calculate quantum capacitance
effects by coupling the Poisson equation with a
Thomas-Fermi approach for the electronic structure. As a first test,
we apply our formalism to analytically solvable models.\cite{citeulike:11024917}

Consider the electrostatic problem of the capacitive coupling between
two conductors at different voltages, separated by a medium with
permeability $\varepsilon(\rr)$.  Classically, the resulting
electrostatic potential $\Phi(\rr)$ follows the Poisson equation
\begin{equation}
\label{poisson-equation}
\nabla\cdot(\varepsilon(\rr)\nabla\Phi(\rr))=-\frac{\rhoF(\rr)}{\varepsilon_0},
\end{equation}
where $\rhoF(\rr)$ is the free charge density and $\varepsilon(\rr)$ is the
relative permittivity. A given potential difference $\ddd \Phi$
between the two conductors $\mathcal{C}_i$, $i=1,2$ will lead to a
charge accumulation 
\begin{equation}
\ddd Q=\int_{\mathcal{C}_1}\ddd\rhoF(\rr)\mathrm{d}\rr - \int_{\mathcal{C}_2}\ddd\rhoF(\rr)\mathrm{d}\rr
\end{equation}
according to Eq.~\ref{poisson-equation}. The classical capacitance $C_{\mathrm{cl}}$
then gives the ratio between charge and potential difference,
\begin{equation}
C_{\mathrm{cl}} = \frac{\ddd Q}{ \ddd \Phi}.
\end{equation}
For nanoscale devices, an additional contribution to the capacitance
can arise due to the electronic structure near the Fermi energy of the
conductor. This quantum capacitance contribution is related to the
additional energy cost for adding electrons to the conductor, which
increases if the DOS near the Fermi edge decreases. Graphene
with its vanishing DOS at the Dirac point is the prototypical point in
case.

Within a Thomas-Fermi approach the local electron density $n(\rr)$
related to the net free charge density through $n(\rr) = \rhoF(\rr)/e$ is 
given by
\begin{equation}\label{eq:nr}
n(\rr) = \int_0^{\EFr} D(E;\rr) \;\mathrm{d}E,
\end{equation}
with $D(E;\rr)$ the local density of states, and $\EFr$ the local Fermi
energy.  In line with the semiclassical limit underlying the
Thomas-Fermi approximation we consider variations only over length
scales large compared to the de Broglie wavelength of the electrons.
Accordingly, effects such as size quantization features of the
graphene nanostructures are neglected in Eq.~(\ref{eq:nr}). They could
be incorporated, e.g., through a fully self-consistent solution of the
Poisson equation and the mean-field Schr\"odinger equation. However,
for the relatively large size of the nanoribbons and in the presence
of a small degree of disorder, the deviation from the Thomas-Fermi
limit is expected to be small. The leading quantum correction to the
classical capacitance should be captured by Eq.~(\ref{eq:nr}). Accordingly, the
electrochemical potential $\mu(\rr)$ contains in addition to the electrostatic
potential $\Phi(\rr)$ the contribution from the local Fermi energy $\EFr$,
\begin{equation}\label{eq:muphi}
\mu(\rr) = \Phi(\rr) + \EFr.
\end{equation}
The solution of the coupled system of Eqs.~(\ref{poisson-equation}),
(\ref{eq:nr}), and (\ref{eq:muphi}) allows the calculation of quantum
capacitance effects. 

Integrating the charge density over the spatial coordinates yields the
total capacitance that relates the accumulated charge with the applied
potential difference. The total inverse capacitance
$C_{\mathrm{tot}}^{-1}$ is given by 
\begin{equation}\label{eq:Ctot}
C_{\mathrm{tot}}^{-1}=\frac{\ddd\mu}{\ddd Q} =
\frac{\ddd\Phi}{\ddd Q} + \frac{\ddd \EF}{\ddd Q} = C_{\mathrm{cl}}^{-1} +
C_{\mathrm{qm}}^{-1}.
\end{equation}
where the inverse quantum capacitance reads 
\begin{equation}\label{eq:Cqm}
C_{\mathrm{qm}}^{-1} = \frac{\ddd\EF}{\ddd Q}= \frac{1}{e D(\EF)},
\end{equation}
and is inversely proportional to the density of states at the Fermi
level. For the latter we use in the following the bulk limit in line
with the Thomas-Fermi approximation, while the variation of $\mu(\rr)$ and
$\EFr$ over the length scale of the device are fully included. The
relative importance of quantum capacitance corrections is governed by
the ratio
\begin{equation}\label{eq:ratioCC}
  \frac{C_\mathrm{{qm}}^{-1}}{C_{\mathrm{cl}}^{-1}}=\frac{\ddd\EF}{\ddd\Phi}=\frac{1}{eD(\EF)}\frac{\ddd Q}{\ddd\Phi},
\end{equation}
i.e.~the ratio of the electrostatically induced charge on the
capacitor, $\ddd Q$, to the total charge $eD(\EF)\ddd\Phi$ near the Fermi edge
accessible by a potential difference $\ddd\Phi$. In the classical
limit, the charge induced on a capacitor at finite voltage is
small compared to the total number of electrons at the Fermi
level. Conversely, a small  density of states at the Fermi level
implies a large $1/C_{\mathrm{qm}}$, and thus a reduction of the
total capacitance in Eq.~(\ref{eq:Ctot}).

\begin{figure}
\includegraphics[draft=false,keepaspectratio=true,clip]{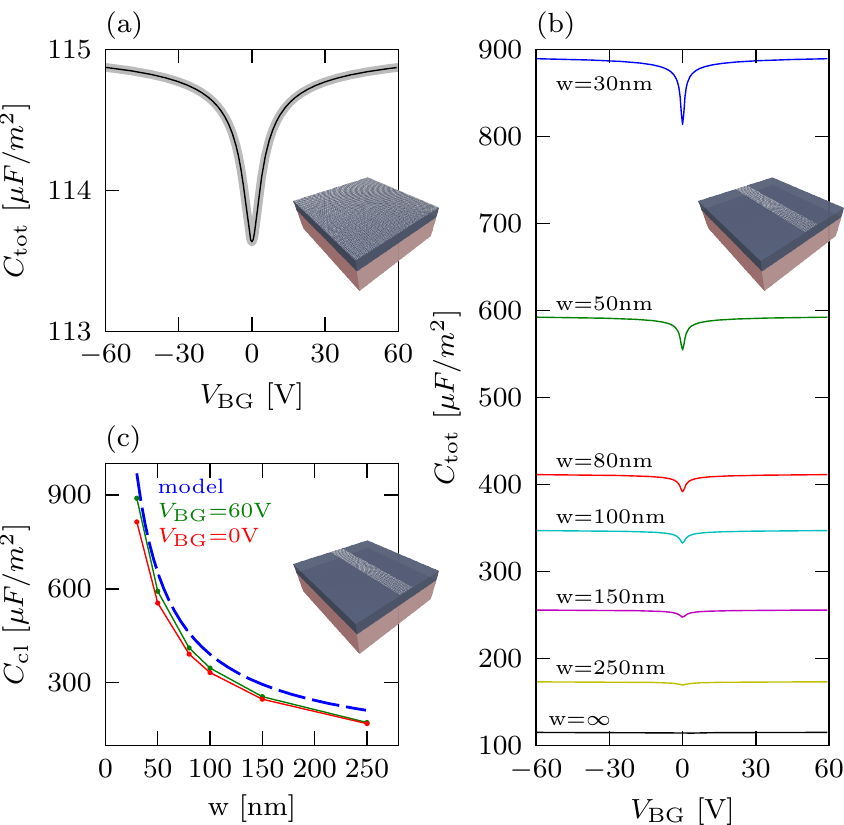}%
\caption{\label{noside-gates} (a) Capacitance of an infinite graphene
  sheet with respect to a back gate with $d=300$ nm, $\varepsilon=3.9$
  and $T=300K$. The analytic solution~\cite{citeulike:11024917} (gray
  line) coincides exactly with our numerical calculation (black
  line). (b) Capacitance of a graphene nanoribbon as a function of
  back gate voltage for different nanoribbon widths $w$ (see
  insets). $w=\infty$ denotes the bulk limit shown in (a). For
  narrower ribbons, the quantum capacitance dip at
  $V_{\mathrm{BG}}=0$~V becomes increasingly prominent. (c)
  Width-dependent capacitance of graphene nanoribbon: analytical
  model (from electrostatics, blue dashed line) and at $V_{\mathrm{BG}}=0$
  (quantum capacitance, red trace) and $V_{\mathrm{BG}}=60$ (classical limit, green trace).  }%
\end{figure}

\begin{figure}[t]
\includegraphics[draft=false,keepaspectratio=true,clip]{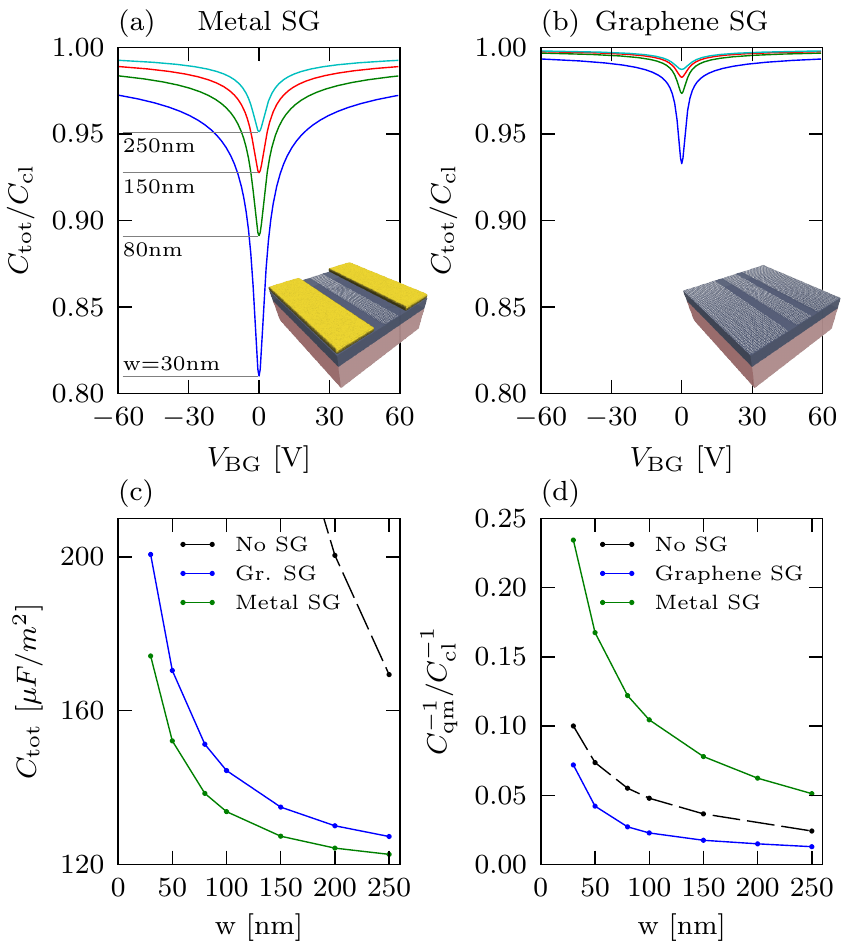}%
\caption{\label{grounded_sg}Voltage-dependent relative change in
  capacitance (relative to respective classical capacitance
  $C_{\mathrm{cl}}$) of graphene nanoribbons featuring (a) grounded
  metal side gates and (b) grounded graphene side gates, for different
  nanoribbon widths [Same parameters as in
    Fig.~\ref{noside-gates}(a)]. Width dependence of (c) the total
  capacitance and (d) the ratio of classical and quantum capacitance
  [see Eq.~(\ref{eq:ratioCC})] for a nanoribbon with no side gates
  (analytical model from \onlinecite{citeulike:11024917}, dashed black
  line), graphene side gates (blue solid line), and metal side gates
  (green solid line).} %
\end{figure}

For infinitely extended two-dimensional graphene, $\EFr$ is the
energy difference between the highest occupied state on the Dirac cone
and the charge neutrality point (i.e.~the so-called Dirac point), as a
function of the free charge carrier density. We can insert the linear
DOS of the Dirac cone, yielding
\begin{equation}\label{eq:Cqmgraph}
\rhoF(\rr)=\frac{\EFr}{\pi(v_{\mathrm{F}}\hbar)^2}
\end{equation}
and
\begin{equation}
  C_{\mathrm{qm}}^{-1}=\frac{\pi(v_{\mathrm{F}}\hbar)^2}{e\EF}.
\end{equation}
We note that edge roughness or
dopants in realistic finite-size structures may strongly
influence device properties, in particular the capacitive coupling
between nanoribbon and side gates.

Finally, we consider finite temperature effects. Since the energy
scales associated with variations of the DOS are of the order of
thermal energies, such corrections are important. We therefore
investigate the relevance of quantum capacitance corrections at, e.g.,
room temperature. At finite temperature $T$, the occupation of
electronic states is smeared out by the Fermi-Dirac distribution
function $f(E,T)$ modifying the expression for the density
[Eq.~(\ref{eq:nr})] to
\begin{equation}\label{eq:nrT}
n(\rr; T) =\int_{0}^{\infty}D(E)f(E - \EFr, T)\;\mathrm{d}E.
\end{equation}
The total charge carrier density (electron and hole charge) for the Dirac cones
in graphene then becomes
\begin{equation}
\label{temperature-rho-e}
\rhoF(T)=\frac{2e}{\pi}\left(\frac{k_B T}{\hbar v_F}\right)^2\left[\mathrm{Li}_2(-e^{\eta})-\mathrm{Li}_2(-e^{-\eta})\right],
\end{equation}
where $\mathrm{Li}_n(x)$ is the polylogarithm and $\eta=E_F/k_B T$.
Inverting Eqs.~(\ref{eq:nrT}) and (\ref{temperature-rho-e}) yields
$\EFr$, which can then be used to solve Eq.~(\ref{eq:muphi}). All data
we present in the following are evaluated at room temperature ($T=300$~K).


\begin{figure*}[t]
\includegraphics[draft=false,keepaspectratio=true,clip]{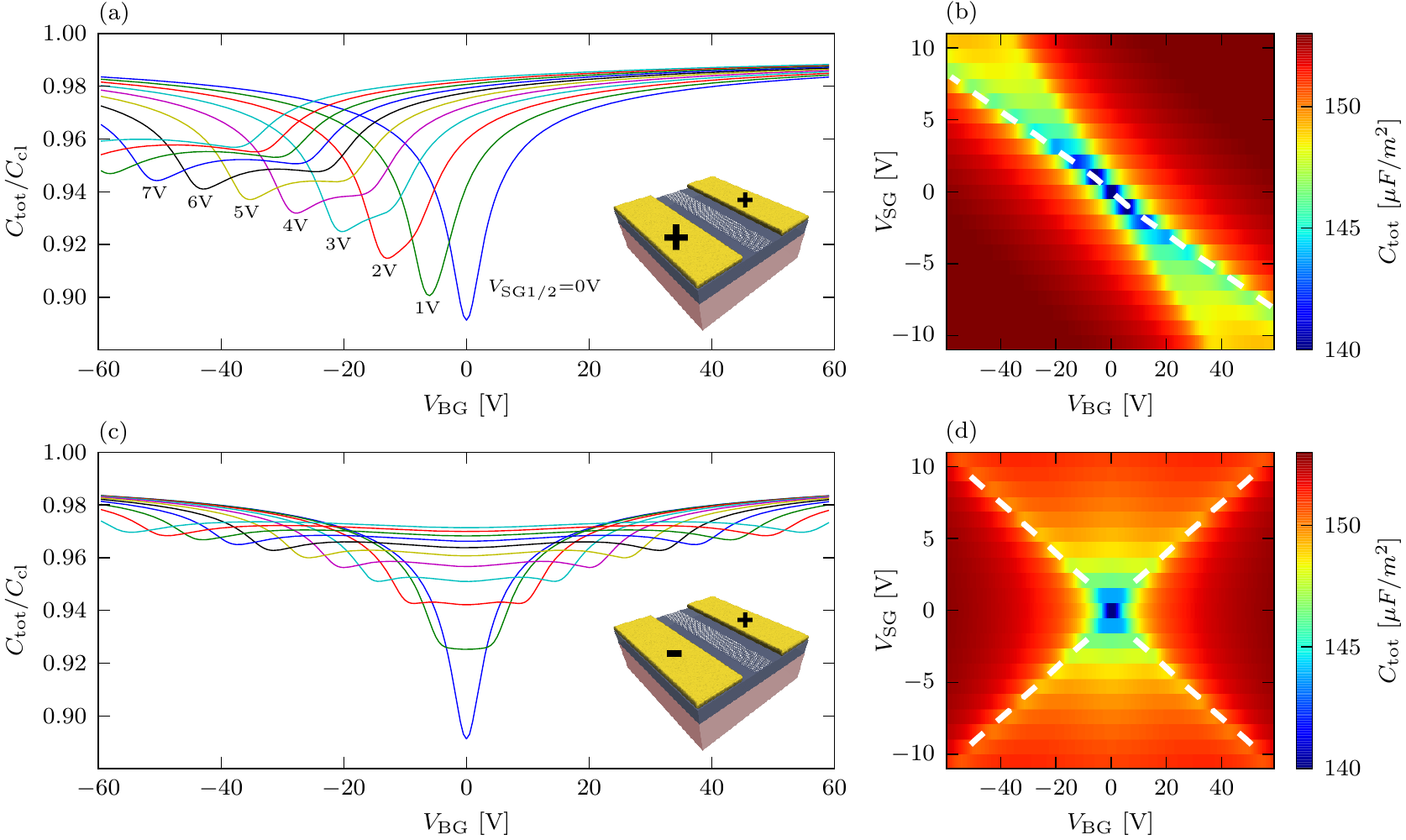}%
\caption{\label{metal_finiteV} Capacitance of a graphene nanoribbon
  with metallic side gates at symmetric [$V_{\mathrm{SG1}}=
    V_{\mathrm{SG2}}$, top row panels (a),(b)] and antisymmetric
  [$V_{\mathrm{SG1}}= -V_{\mathrm{SG2}}$, bottom row panels (c),(d)]
  finite side gate voltages (see insets). (b),(d) Contour plots of the
  total capacitance as a function of $V_{\mathrm{SG}}$ and
  $V_{\mathrm{BG}}$. Dashed white lines mark the voltage combinations
  where charge accumulated at (one of the) edges of the nanoribbon is
  zero.}
\end{figure*}

\section{Results and discussion}\label{sec:results}

We aim to describe a capacitor formed by a graphene nanoribbon, a
metallic back gate and two side gates [see
  Fig.~\ref{system-illustration}(a)]. The step from idealized,
infinite graphene to a nanoribbon with side gates (of possibly
different materials and at different potentials) introduces several
new device-specific quantum capacitance-related effects. We
disentangle them by considering configurations of increasing
complexity. To test the validity of our numerical simulation, we start
our discussion with the simplest case of an infinitely extended
graphene sheet [see Fig.~\ref{noside-gates}(a)]. The nanocapacitor is
completed by an infinite back gate at a distance of $d=300$ nm~[see
  Fig.~1(a)], the gap between the two sheets filled with a dielectric
substrate, $\mathrm{SiO_2}$ with $\varepsilon=3.9$; above the
nanoribbon is air. The boundary conditions for $\mu(\rr)$ are given by
the potentials applied externally to the different gates. We assume translational
symmetry perpendicular to the cross-section shown in
Fig.~\ref{system-illustration}(b), thus reducing the calculation to
2D (For technical details, see App.~\ref{App:technical}). As
expected, the capacitance decreases at the Dirac point [see
  Fig.~2(a)] signifying the quantum capacitance effect due to the
reduction of the DOS at the Dirac point. Our
numerical results agrees (within the numerical accuracy) with
analytical models \cite{citeulike:11024917}.

\begin{figure*}[t]
\includegraphics[draft=false,keepaspectratio=true,clip]{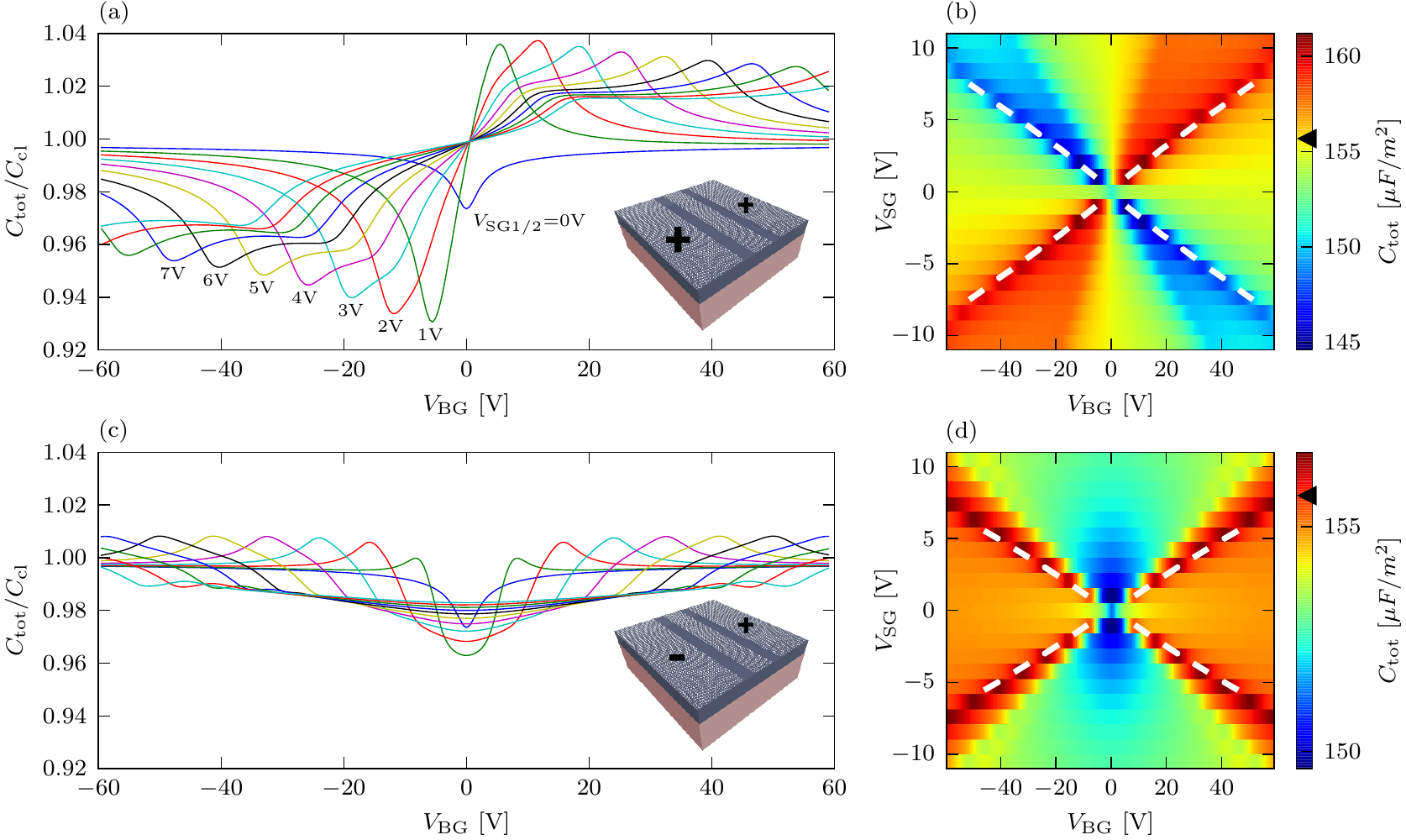}%
\caption{\label{graphene_finiteV} Same as Fig.~\ref{metal_finiteV}
for graphene side gates. Note the regions of increased capacitance
due to quantum-capacitance effects in the side gates (blue areas in color maps).
These regions are delimited by voltage combinations where
charge vanishes at the ribbon edges (white dashed lines). Black triangles
on the color scale denote the classical capacitance value. 
}%
\end{figure*}

\subsection{Graphene nanoribbons without side gates}

For a nanoribbon of finite width $w$ but without side gates, the
quantum capacitance effect is substantially increased due to the
smaller DOS, leading to a more pronounced quantum capacitance dip for
narrower nanoribbons [see Fig.~2(b)]. Consequently, an inclusion of
quantum capacitance effects becomes more important for smaller
nanonstructures. Concurrently, the classical capacitance per unit area
of a nanoribbon increases with decreasing nanoribbon width [see
  Fig.~2(c)], further amplifying the absolute change in capacitance at
the Dirac point. We find good agreement between our simulation and the
analytical model discussed by Lin et al.~\cite{lin08} [see blue dashed
  line in Fig.~2(c)],
\begin{equation}
C(w)=C_{\mathrm{cl}} \left[ \frac{2}{\pi} \mathrm{arctan}\left( \frac{w}{4d} \right)
+ \frac{w}{4d \pi} \mathrm{ln}\left( 1+ \frac{16 d^2}{w^2} \right)  \right]^{-1}. \nonumber
\end{equation}
 The residual differences are most likely due to the use of spatially
 constant effective dielectric constants in the analytical model. As
 discussed above, one can use a full quantum local density of states
 (LDOS) in Eq.~(\ref{eq:nr}) beyond the Thomas-Fermi
 approximation. Using the quantum LDOS calculated for this geometry we
 have verified that such a correction yields a negligible change in
 the capacitance for realistic nanoribbon widths. In line with
 Eq.~(\ref{eq:ratioCC}) we expect the approach of the classical limit
 with increasing width $w$ [Fig.~2(b)] illustrating that $
 C_\mathrm{{qm}}^{-1} / C_{\mathrm{cl}}^{-1}\rightarrow 0$ as
 $w\rightarrow\infty$ [Fig.~3(d)].

\subsection{Graphene nanoribbons with side gates}

The introduction of side gates precludes an analytic solution of the
problem. Side gates decrease the overall capacitance of the
nanoribbon since the classical field-lines further away from the
nanoribbon no longer bend towards the nanoribbon (increasing its
capacitance), but towards the side gates [see Fig.~1(b)]. Therefore,
charge will accumulate at the side gates which decreases the
nanoribbon capacitance [compare scale on Figs.~3(c) and 2(c)]. In
addition to this classical effect, the different DOS in the graphene
nanoribbon and metal side gates further increase the quantum
capacitance effect [see Fig.~3(a)]. Put simply, it is energetically
much more costly to put electrons into the nanoribbon (due to its
small size and the small DOS) than into the metal side gates. A
drastically different behavior emerges for graphene side gates [see
  Figs.~3(b),(c) and (d)], since at the Dirac point the DOS decreases
in both the side gates and the nanoribbon simultaneously.

In the experiment, the electrostatic tuning of device properties
proceeds by varying the voltages of the side gates relative to that of
the back gate. As two prototypical examples, we treat the symmetric
(i.e.~$V_{\mathrm{SG1}}=V_{\mathrm{SG2}}$) and antisymmetric
(i.e.~$V_{\mathrm{SG1}}=-V_{\mathrm{SG2}}$) voltage configurations.
Due to the non-uniform potential distribution $\Phi(\rr)$, different
parts of the device reach the (local) Dirac point at different voltage
configurations, namely when locally $\mu(\rr)=\Phi(\rr)$. 

We first consider metal side gates and the symmetric voltage
configuration: for positive side gate voltages, the position of the
quantum capacitance dip shifts to negative back gate voltages where
the side gate influence is compensated [see Figs.~4(a) and 4(b)].
When the energy of the Dirac point is reached in the nanoribbon, the
capacitance decreases. The relative depth of the dip depends inversely
on the nanoribbon width. For finite voltages $V_{\mathrm{SG}}$, the side
gates cause a nonuniform electrostatic potential in the nanoribbon,
and the Dirac point is thus reached at different positions in the
nanoribbon for different back gate voltages. With increasing side gate
voltage, the non-uniformity of the electrostatic potential
grows. Consequently, the quantum capacitance dip becomes wider and
more shallow: a capacitance landscape in the
$V_{\mathrm{BG}}$-$V_{\mathrm{SG}}$ plane reveals two lines crossing
each other at a small angle [see white lines in Fig.~4(b)] denoting
the voltage combinations where no local charge is induced by the gates
at either the center of the nanoribbon or its edges. These
combinations control the width and shape of the quantum capacitance
dip. For the antisymmetric voltage combinations, the situation changes: 
the width of the dip is now given by the voltage combinations where there
is no charge on either the left or the right edge of the nanoribbon
[see white lines in Fig.~4(d)], making the change in dip
shape more obvious [see Fig.~4(c)].


For graphene side gates, the situation becomes more complicated and
novel features emerge: since the DOS of both the nanoribbon and the
side gates features minima at the respective Dirac points, 
quantum capacitance (QC) effects in the nanoribbon
and the side gates interact, leading to features beyond the standard
``QC dip''. We find an increase in capacitance for larger side gate
voltages [i.e.~a quantum capacitance peak, see Fig.~5(a)]. This
counter-intuitive \emph{increase} in capacitance due to the limited
density of states rather than a reduction arises from the reduced
screening response by the side gates which, in turn, restores the
capacitance of the nanoribbon.  The screening ability of graphene side
gates suffers when the Dirac point is reached within the side
gates. This loss in screening thus causes a positive peak in the
nanoribbon capacitance [see top right in
  Fig.~\ref{graphene_finiteV}(a)].  These results imply the surprising
 phenomenon of an effectively negative quantum
capacitance. Previously, (narrow) negative quantum capacitance peaks
were only observed for graphene heavily doped with Ag adatoms, leading
to dispersionless resonant impurity bands near the charge neutrality
point \cite{Wang13}.  The explanation for the splitting and
broadening of the QC dips applies analogously to the QC peaks [see
  dashed lines in Figs.~5(c),(d)].

A comparison of the size of the quantum capacitance dip suggests that
the quantum capacitance effect is larger for grounded
($V_{\mathrm{SG1}}=V_{\mathrm{SG2}}=0$) metal than grounded graphene
side gates [compare Figs.~4(a) and 4(b)]. The observation that
the capacitance of the nanoribbon increases when the side gates locally
reach their Dirac point offers an additional explanation why the QC
dip is deeper for metal side gates than for graphene side gates: since
the nanoribbon and the side gates reach the Dirac point at the same
potential, the large negative nanoribbon QC dip is 
superimposed on a positive effective side gate QC peak, and thus,
the dip depth is decreased. This overlap can be seen 
in the capacitance distribution in the $V_{\mathrm{BG}}$-$V_{\mathrm{SG}}$
plane [see Figs.~5(c),(d)]: regions of low (blue) and high (red)
capacitance converge at the center.  As soon as the graphene side
gates are at a finite electrostatic potential, the QC peak and the QC
dip are shifted relative to each other causing the
broadening of the peak and an increase in dip depth.

\begin{figure}[t]
\includegraphics[draft=false,keepaspectratio=true,clip]{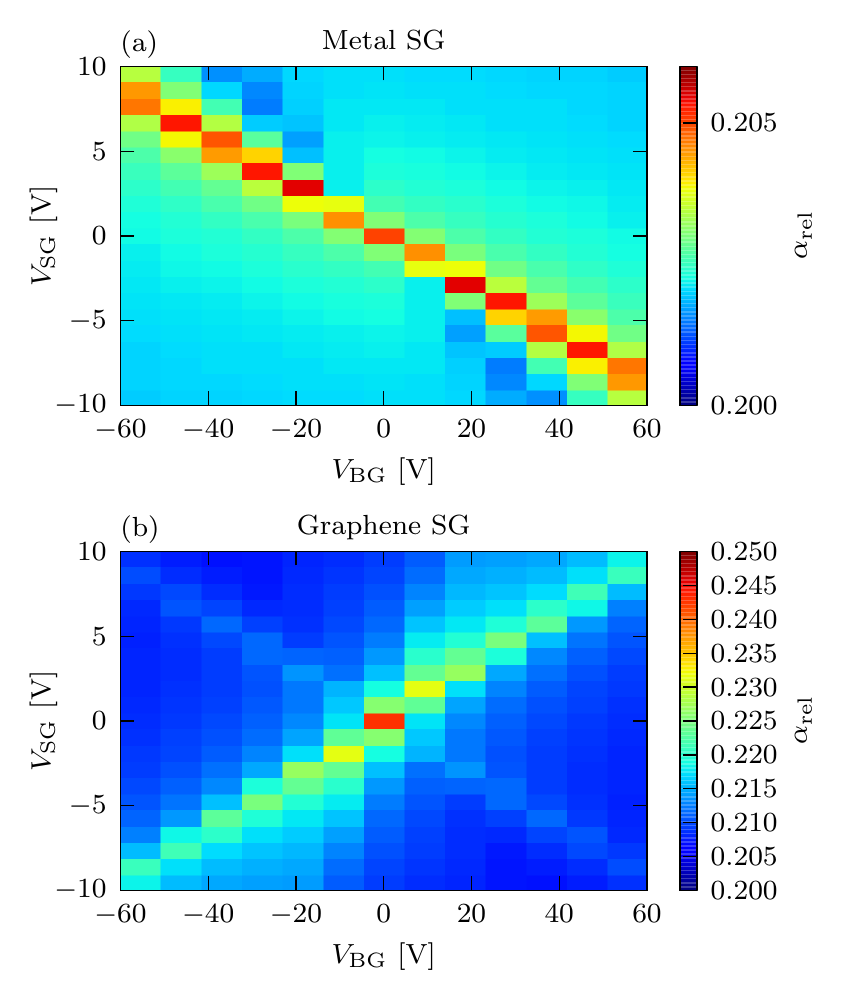}%
\caption{\label{fig:lever} Relative lever arm $\alpha_{\mathrm{rel}}$
  [see Eq.~(\ref{eq:alpha})], i.e., relative capacitance of the
  nanoribbon towards back gate and side gate, in symmetric gate
  voltage configuration for (a) metallic and (b) graphene side
  gates. The graphene nanoribbon has a width of 80 nm and the spacing
  to the side gates is 30 nm. }
\end{figure}

\subsection{Relative lever arms}

Instead of considering the back gate capacitance as modified by
side gates, one may alternatively consider separately the charges induced on the
nanoribbon due to finite side gate and the back gate voltages: for
each gate, we define a capacitance relating the voltage on the
gate to the induced charge on the nanoribbon. A comparison
of the relative capacitive couplings, i.e.~the ratios of the
capacitances, is of great interest for experiments: back gate and side
gates usually feature very different coupling coefficients due to the
different spatial distance to the nanoribbon. This
relative coupling is usually referred to as relative lever arm
\begin{equation}\label{eq:alpha}
\alpha_{\mathrm{rel}} = C_{\mathrm{BG}}/C_{\mathrm{SG}}.
\end{equation}
Accurate knowledge of $\alpha_{\mathrm{rel}}$ is essential for a
detailed interpretation of experimental results. Due to quantum
capacitance effects, $\alpha_{\mathrm{rel}}$ depends on
$V_{\mathrm{BG}}$ and $V_{\mathrm{SG}}$, as we now investigate in
detail.  We determine side gate capacitance $C_{\mathrm{SG}}$ as
$C_{\mathrm{SG}}=\left.\ddd Q_{\mathrm{GNR}}/\ddd
V_{\mathrm{SG}}\right|_{V_{\mathrm{BG}}}$ and
$C_{\mathrm{BG}}=\left.\ddd Q_{\mathrm{GNR}}/\ddd
V_{\mathrm{BG}}\right|_{V_{\mathrm{SG}}}$.  The relative lever arm as
function of $V_{\mathrm{BG}}$ and $V_{\mathrm{SG}}$ features a
substantial energy dependence already for metallic side gates [see
  Fig.~\ref{fig:lever}(a)]: we find the relative lever arm decreased
by roughly three percent around zero carrier density in the nanoribbon. For the
case of graphene side gates, we observe a decrease by up to fifteen
percent of the relative lever arm along a positive diagonal in the
gate voltage plane [highlighted in Fig.~6(b)]. This is of importance
as experimental devices often feature side gates and nanoribbon etched
from the same graphene sheet.


Interestingly, when comparing our numerical results with the
experimental data reported in Ref.~\onlinecite{mol09} one finds that the
experimentally extracted $\alpha_{\mathrm{rel}}$ is nearly a factor three
larger than our calculation suggests. Since the back gate coupling is
well controlled in experiments on large-scale graphene devices, we
conjecture that this discrepancy might be related to a significantly
reduced side gate coupling in the experiment. In particular, the edges
of a realistic experimental nanoribbon device feature edge roughness
and uncontrolled edge terminations (both not included in the present
model) which may lead to a large local density of states at the edges,
and thus to a significant charge accumulation at the edge of the
nanoribbon. These charges may partly screen the lateral side gates and
therefore increase the measured relative lever arm.  Such a screening
effect is also consistent with recent observations on hydrofluoric
acid (HF) treated graphene nanoribbons where an increased side gate
coupling strength has been found after HF dipping~\cite{dau13}.

\section{Summary and outlook}

We have shown that side gated graphene nanoribbons exhibit significant
variation in capacitance as a function of gate voltages due to
classical screening and quantum capacitance effects in the
nanoribbon. Both the nanoribbon geometry as well as the presence and
material of side gates strongly influence quantum capacitance
effects. We find both positive and, surprisingly, unconventional
negative quantum capacitance corrections. The former is the usual
decrease of total capacitance due to the additional energy required to
fill electronic states at a low density of states. The latter occurs
due to a decrease of screening by graphene side gates. Since the
capacitance of a graphene nanodevice frequently enters in the tuning
of the effective energy of electrons in, e.g., transport experiments,
inclusion of quantum capacitance effects is critical for
the correct interpretation of experimental data. The proposed
increase in total capacitance for finite voltage at graphene side
gates and the significant change in the relative gate lever arms
should be observable experimentally for clean samples.

\section*{Acknowledgment}

Financial support by the DFG (SPP-1459 and FOR-912), the ERC, the FWF
SFB ViCoM, and the JARA Seed fund is gratefully acknowledged.

\appendix

\section{Numerical method}\label{App:technical}

In the following, we provide further technical details on our
calculations. We solve the Poisson equation (\ref{poisson-equation})
using a finite difference scheme with a grid spacing of 1
nm. Dirichlet boundary conditions are applied to the conducting parts
of the system: the lower boundary (where the back gate fixes the
potential), the nanoribbon and the side gates. The left, right and top
border are best modeled using Neumann boundary conditions,
i.e. $\frac{d\Phi}{d\mathbf{n}}=0$, as these emulate the infinitely
extended system better than Dirichlet boundary conditions. We solve
Eqs.~(\ref{eq:muphi}) for $\Phi(\rr)$ using the charge density for
graphene, Eq.~(\ref{eq:Cqmgraph}), in combination with the Poisson
equation (\ref{poisson-equation}). After the discretization, the
resulting nonlinear system of equations can be solved using any
nonlinear solver. On all grid points $\rr_i$ of the graphene parts,
there is a difference between the electrochemical potential
$\mu(\rr_i)$ and the electrostatic potential $\Phi(\rr_i)$, given 
by a discretized version of Eq.~(\ref{eq:muphi}). Inserting these
constraints into the Poisson equation (\ref{poisson-equation}) 
yields
\begin{equation}
\label{qcinfgraphene-charge-grid}
\varepsilon_0\Delta\Phi(\rr) = \left\{
\begin{array}{ccl}
 \frac{q}{\pi}\left(\frac{\mu(\rr_i)-\Phi(\rr_i)}{v_F \hbar}\right)^2
&,& \rr \in \rr_i\\
0&,&\rr \notin \rr_i
\end{array}
\right.
\end{equation}
where $\rr_i$ are the grid points on the graphene parts. This system of
equations ($\rr = \rr_i$, $i=1,\ldots,N$) is solved for $\Phi(\rr_i)$. 

Quantum capacitance can also be formulated for a non-local relationship
between energy and charge (as required, e.g., for ab-initio calculations
of the system)
\begin{equation}
\label{qcequation-energy-app}
\mu(\rr)=\Phi(\rr)+E[\varepsilon(\rr)\varepsilon_0\Delta\Phi](\rr).
\end{equation}
This equation can be reformulated in terms of charge
densities instead of energies by applying the inverse map
$\EFr^{-1}=\rho_q[\Phi(\rr,\mu(\rr))]$,
\begin{equation}
\label{qcequation-charge-app}
\rho_q[\mu,\Phi](\rr)=\varepsilon_0\Delta\Phi(\rr).
\end{equation}
which gives the charge density, $\rhoF$, as a function of the
potential $\Phi(\rr)$ and the chemical potential $\mu(\rr)$. Assuming
a local dependence between $\EFr[\rhoF]$ and $\rhoF$ allows replacing
the functional formulation (\ref{qcequation-energy-app}) by the simpler
(\ref{eq:muphi}) used in the main text.

\end{document}